\documentclass[vecphys]{svmult}

\usepackage{graphicx}        
\usepackage[bottom]{footmisc}

\begin{document}
\title*{Near-infrared VLT adaptive optics imaging of evolved stars}
\author{Eric Lagadec\inst{1} \and
Olivier Chesneau\inst{2} \and
Albert A. Zijlstra\inst{1}\and
Mikako Matsuura\inst{3} \and
Djamel M\'ekarnia\inst{2}}
\authorrunning{Lagadec et al.}
\institute{University of Manchester, School of Physics and Astronomy, M60 1QD Manchester (UK)
\texttt{eric.lagadec@manchester.ac.uk}
\and  Observatoire de la C\^ote d'Azur (France)
\and NAOJ (Japan)}
%
%
\maketitle

\begin{abstract}
The high angular resolution and dynamic range achieved by the NACO adaptive optics system on the VLT is an excellent tool 
to study the morphology of Planetary Nebulae (PNe). We observed four stars in different evolutionary stages from the
 AGB to the PNe phase.  The images of the inner parts of 
the PN Hen\,2-113 reveal the presence of a dusty torus tilted with respect to all the other structures of the nebula.
 A void
 0.3 arcsec in diameter was discovered with NACO around the central source. These data indicate the presence of a
 cocoon of hot dust (T$\sim$1\,000 K) with a total mass ~10$^{-9}$M$_{\odot}$ in the core of the nebula. This was 
not expected
 so close to a hot (T$_{eff}$$\sim$30\,000K) central star, and our observations indicate that dust is present
 close to this 
central star.
The NACO observations of Roberts\,22 reveal an amazingly complex nebular morphology with a S-shape that  can be
 interpreted in terms of the 'warped disc' scenario of Icke (2003). Comparing these observations with previous HST 
images seems to indicate that this S-shaped structure is expanding at $\sim$450 km.s$^{-1}$. Combined NACO and MIDI 
(the VLTI mid-infrared
 interferometer) observations of the nebula OH\,231.8+4.2 have enabled us to resolve a very compact (diameter of 30-40 mas
, corresponding to 40-50 a.u.)
 dusty structure in the core of the nebula. Finally, recent observations of the AGB star V Hydrae show that this
 star present a departure from spherical symmetry in its inner shell and is probably on its way to become an asymmetrical
 planetary nebula.
These observations show that NACO is a great
 instrument for the discovery and study of small structures in circumstellar envelopes and PNe and a good complement to interferometric devices. 
\end{abstract}

\section{Introduction}
High sensitivity and angular resolution observations are of high interest for the understanding of PNe
 morphologies.
Thus, the appearance of CCD detectors in the 80s and then the successful launching of the Hubble Space Telescope (HST)
 have
 helped to reveal an amazingly rich variety of PNe morphologies (see Balick and Frank (2002)). With the installation of adaptive optics instruments on 8m-class telescopes, it is now
 possible to obtain good sensitivity images from the ground thanks to real time correction of the atmospheric 
turbulence. Combining this with the angular resolution achieved with such telescopes provides a great tool for the 
study of very small structures in the heart of PNe, as well as circumstellar envelopes around their 
precursors AGB stars and pre-PNe (PPN).

We used the adaptive optics imager NACO on the VLT, which can provide diffraction limited images in the near-infrared
 and spatial resolution around 60\,mas in K and L bands, to  observe 4 objects in different evolutionary phases from 
the AGB to the PNe stage. These object are surrounded by dust so that infrared imaging allows us to 
deeply study inner details of their envelopes or nebula.
In this paper we present the results from our observations of the planetary nebulae Hen\,2-113 and Roberts\,22, the 
nebula OH\,231.8+4.2 and the extreme AGB star V Hya.
%
\section{Hen 2-113}
\begin{figure}
\centering
\includegraphics[height=6cm]{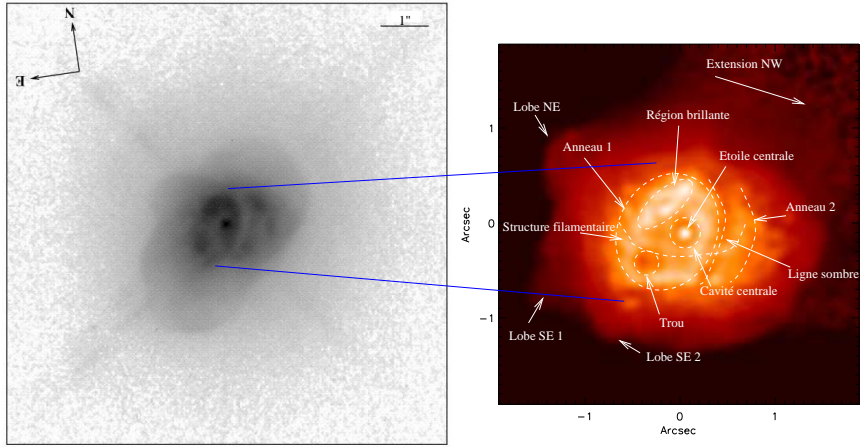}
\caption{Left: optical HST image of Hen 2-113 (Sahai et al. 2000). Right: NACO image of Hen 2-113 at 3.74$\mu$m (Lagadec
 et al. 2006).}
\label{fighen}     
\end{figure}

Hen 2-113 (hereinafter HEN) is a PN with a [WC10] central star.
HST observations of this object by Sahai et al. (2000) have shown that
HEN exhibits a complex geometry, roughly bipolar with two bright,
knotty, compact ring-like structures around the central star (See left panel of Fig. \ref{fighen}). This compact structure is embedded in a larger and fainter
 spherically symmetric AGB envelope remnant.
Infrared observations of HEN were obtained with NACO and MIDI (Lagadec et al. 2006). 
We also attempted to detect and study small scales 
structures in the Mid-IR with the long baseline interferometer MIDI but the nebula appeared over-resolved with 46m
 baselines so that no interferometric data could be recorded.
HEN exhibits a clear 1" torus-like structure
superimposed to a more diffuse environment visible in the L' (3.8$\mu$m), M' (4.8$\mu$m) and
8.7$\mu$m bands. We interpret the two ring-like structures as due to the projection of the lobes of a diabolo-shaped
 structure observed with an inclination of about 40$^{\rm o}$.

Photometry of the central star  in $\rm L^\prime$ and $\rm M^\prime$ band indicates that it is $\sim$300 and $\sim$800
times brighter than predicted by stellar models. Moreover, the central object appears resolved in L' band with measured
FWHM of about 155\,mas. Simple calculations indicate that this infrared excess can be
explained by emission from hot dust grains. A mass of $\sim10^{-9}\rm
M_{\odot}$ with T$\sim$900K can account for this infrared excess. By a totally
independent way (fitting of the SED), Sahai et al. (2000) also indicated the
possible presence of hot dust (T$\sim$900K and  $\rm M\sim 10^{-9}\rm
M_{\odot}$) inside the nebula.
This is a clear evidence that hot dust is present real close to the hot [WC] star. Finding hot dust around a hot star (T$_{eff}$$\sim$30\,000K) with a strong wind is surprising and we can wonder whether a single star alone is able to produce this dust in such conditions.  Such dusty structures could be common to PPNe and PNe  as it has been shown that colour measurements in other PNe observed
 with adaptive optics on Keck indicate the presence of hot dust close to the CS among a few of them (e.g.
 IRAS 16342--3814 (Sahai et al., 2005) and IRAS 18276--1431 (S\'anchez-Contreras et al.,2007)).

\section{Roberts 22}

Roberts\,22 (IRAS\,10197-5750) is a
bipolar PN, displaying OH maser emission 
at 1612 and 1665\,MHz. The circumstellar envelope expands with 
a velocity of about $20\,\mathrm{km\,s}^{-1}$ (Zijlstra et al. 2001). Optical spectroscopy shows that the
$\mathrm{H}\alpha$ line profile has wings extending up to 
$\pm\,450\,\mathrm{km\,s}^{-1}$, a signature of high-speed outflows.

In this work, we present new observations of this young PN obtained with NACO at 2.12\,$\mu$m.
\begin{figure}
\centering
\includegraphics[height=6cm]{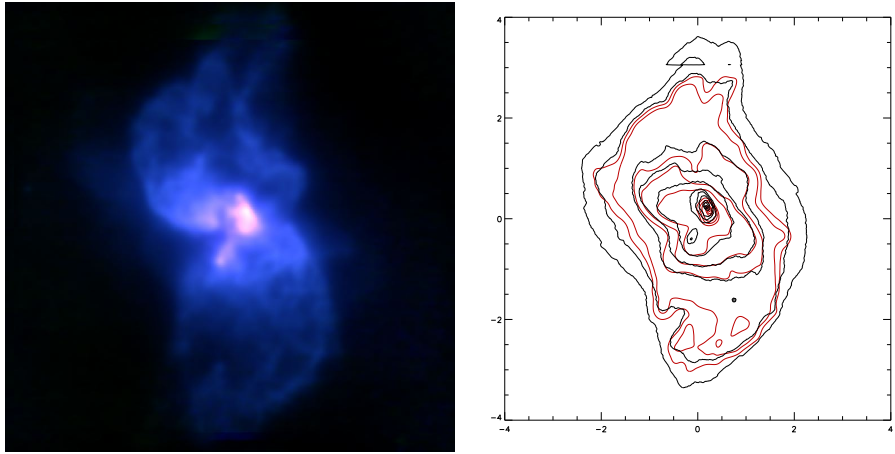}
\caption{Left: NACO image of Roberts\,22 at 2.12\,$\mu$m. Right, contours of Roberts\,22 images at 2.12$\mu$m. Black contours: our NACO observations; red contours: 2003 HST observations.}
\label{figrob}     
\end{figure}
The nebula around Roberts 22 has previously been imaged in the optical by the HST (Sahai et al. 1999). These observations
 show that the nebula is bipolar, with bright lobes shaped like a butterfly's ''wings''. A dark equatorial body due to
 dense dust that obscures the central star is also observed. This dark equatorial structure is oriented at a position angle 
of 120$^{\rm o}$ and is close to, but not exactly parallel to the sides of the parallelogram formed by the bright lobes.
 The bright lobes are surrounded by a faint halo, which morphology is similar to that of the 
bright lobes.

The NACO image of Roberts 22 reveals a complex morphology of its nebula (Lagadec et al, in preparation).
 The parallelogram-shaped structure
 observed by the HST is faintly visible in our NACO images. But inside this structure, the nebula has a complex
 roughly S-shaped morphology. The tail of this S-shaped structure are observed outside of the parallelogram-shaped halo.
 This structure is very clumpy with holes certainly due to a lack of material. These holes are bigger toward the north and
 south edges of the S-shaped structure. The tail of this structures seems to be organised in filaments. Three filaments
 are clearly observed in the north. Two filaments are also observed in the south. A third incomplete one  might be
 present on the South-East.

Some unpublished images of Roberts 22 have been obtained by the HST at 2.12$\mu$m, i.e. at the same wavelength as ours,
 but 3 years before. We thus compared both images to look for evolutions in the shape of the nebula. To make both images 
comparable we had to artificially decrease the spatial resolution of our NACO image by convolving it with a Gaussian
 having the same FWHM as the HST Point Spread Function. We then superimposed the two images using 4 field stars. The
 contours of the two resulting images are shown in Fig.\ref{figrob}. We have to keep in mind that it is very difficult to compare two
 images taken with different instruments as their sensitivities are different, furthermore here, as one image was taken
 from the ground and the other one from space. Nevertheless, these images seems to indicate that the NACO image is more 
extended toward the North-West and South-East (i.e. along the direction of the S-shaped structure) by $\sim$0.1''. At
 the distance of the nebula ($\sim$2kpc), such an extension in 3 years corresponds to an expansion velocity of
 $\sim$450 km.s$^{-1}$, which is the speed of the high velocity outflows measured in H$_{\alpha}$. It is thus
 very tempting to claim that
 the S-shaped structure in the heart of the nebula is expanding at a speed of 450 km.s$^{-1}$ and that the outflow is almost in the plane of the sky. However, to confirm 
this, integral field spectroscopy of Roberts 22 will be needed.  
\newpage
\section{OH\,231.8+4.2}

.

\begin{figure}
\centering
\includegraphics[height=6cm]{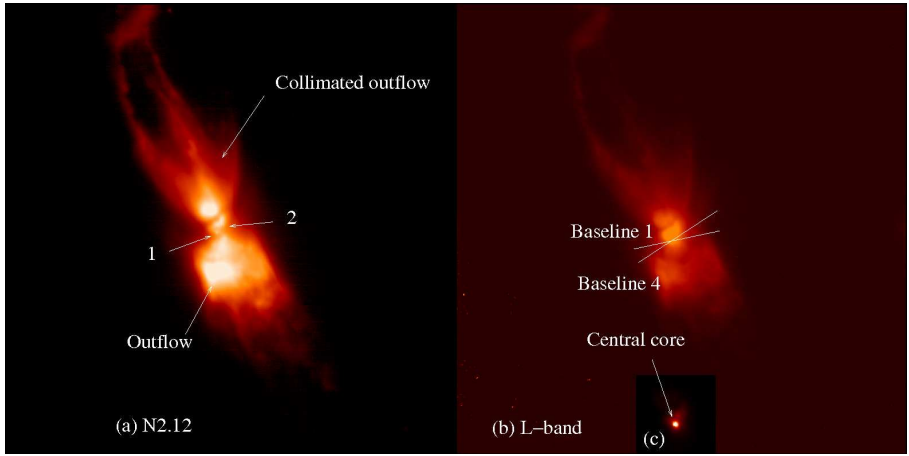}
\caption{Left: NACO image of OH\,231.8 at 2.12$\mu$m. Right NACO image of OH\,231.8 at 3.78$\mu$m. The caption show a
 zoom on the core of the nebula.}
\label{figoh}     
\end{figure}

OH 231.8+4.2 is the nebula around the star QX Pup. It shows two
(ionised) bipolar lobes on either side of a central obscuring lane. The
obscured central star shows Mira-like variability, indicative of an
evolved AGB star. Optical spectroscopy indicates the presence of a hot A-type binary companion to the AGB central star
 (S\'anchez-Contreras et al., 2004).
 We obtained NACO observations of this nebula (Matsuura et al., 2006).
 The resulting data are arranged in Fig. \ref{figoh}. The left panel shows the
2.12 micron image, with the two lobes clearly identifiable.
The right panel, b), shows the L-band image. Panel c) shows
a focus on the core seen on panel b). Here the
central region is shown to contain a bright, very compact, unresolved source. To see if this structure was a dusty disc 
similar to the ones observed at the heart of some PN (see Chesneau et al. and Lykou et al., these proceedings), we
 observed this central source with
 the VLTI mid-infrared interferometer MIDI. This is a two telescope interferometer, using two VLT UT 8.2 telescopes
 for our observations. It thus has a resolution proportional to the distance between the two telescopes in the
 direction formed by the two telescopes (the baseline) and the resolution of a 8.2 meter telescope in the perpendicular
 direction. In a previous work
(Matsuura et al. 2006), we obtained observations using several VLTI/MIDI baselines nearly perpendicular to the observed
 bipolar outflows of the nebula. The unresolved source in our NACO images was resolved, with a diameter of approximately
 30-40 mas.

\begin{figure}
\centering
\includegraphics[height=6cm]{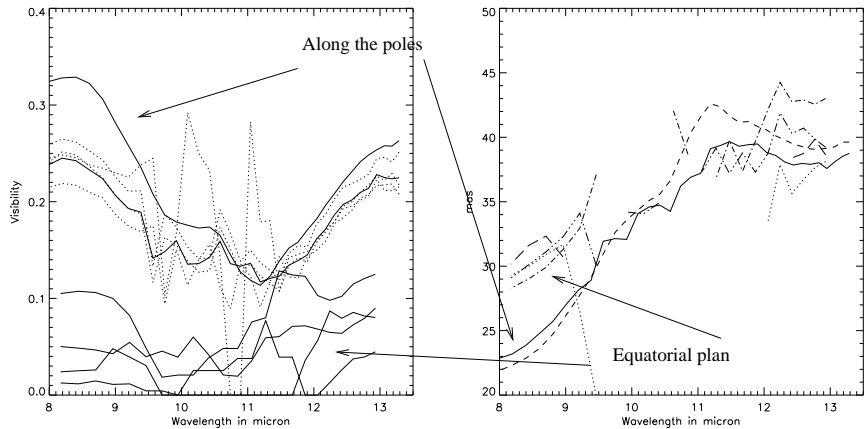}
\caption{Left: MIDI visibilities of OH\,231.8+4.2 for different baselines. Right: Gaussian fit to these visibilities.}
\label{figmidi}     
\end{figure}

However, as our measurements have been only made in one direction, it was impossible to know the shape of  the 
observed compact structure. It could indeed be  a disc or a spherical shell for example. We thus obtained new 
observations with baselines having different orientations along the pole of the nebula. The observed visibilities 
are presented Fig. \ref{figmidi} (for more details on observations with MIDI, we refer to the contribution by
 Chesneau et al., these proceedings).  The Gaussian fit to the visibilities provide a measurements of the size of the 
observed structure
along the baseline direction. The right panel of Fig. \ref{figmidi}
 shows that the shape of the structure depends on the wavelength. At 8\,$\mu$m, the dusty structure is 1.3 times more
 extended along the equator than along the poles of the nebula. At 13\,$\mu$m, this structure has almost
 the same dimension
 in all the directions. These visibilities are difficult to fit in the picture of a nearly edge-on disc. There is a flatten structure at 8$\mu$m that could correspond to the inner rim of te dusty environment, embedded in a more spherical dusty halo (the dense dusty wind of the Mira). Radiative transfer modelling is 
needed
 to confirm this hypothesis.

\section{V Hya}
V\,Hydrae (V\,Hya), is an evolved  cool carbon star located at 380\,pc  from the 
sun (Knapp et al. 1997) with a high mass loss rate. V\,Hya is 
associated  with a very  fast ($>$ 100 kms$^{-1}$) outflow that was 
observed  in the CO J=2-1 and J=3-2 spectra 
(Knapp et al. 1997). Knapp et al. (1997) also proposed that the 
fast-moving gas is expanding along the East-West direction. From recent HST observations,
Sahai et al. (2003) reported the discovery of a
newly ejected high-speed jet-like outflow in this star. These observations,
combined with the previous interferometric CO (J=1-0) map of V\,Hya, favour the
picture of an expanding, tilted and dense disk-like structure, oriented
north-south, present inside the inner envelope.

\begin{figure}
\centering
\includegraphics[height=6cm]{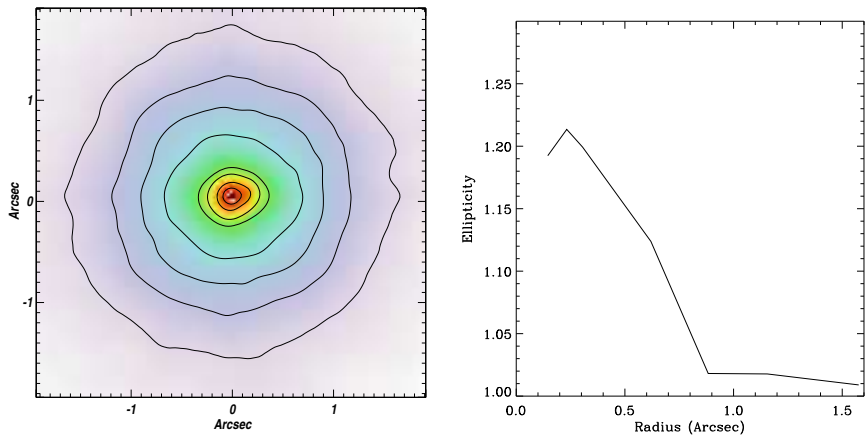}
\caption{Left: NACO image of V\,Hya at 3.74$\mu$m. Right: Ellipticity of the envelope as a function of the radius.}
\label{figvhya}     
\end{figure}

Previous mid-IR images show that the circumstellar envelope 
of V\,Hya is roughly spherically symmetric in the central parts but 
present an elongation in the East-West direction 
at intensity levels lower than 20\,$\%$ of the central intensity peak (Lagadec et al. 2005). 
The observed elongated feature in the East-West direction could be associated 
with the dust emission from material blown away perpendicularly to the
equatorial disk, consistent with the model proposed by Sahai et al. (2003).

We imaged the circumstellar envelope around V Hya using NACO at 3.8$\mu$m  (Fig. \ref{figvhya}, M\'ekarnia et al., in preparation). We clearly see that at large scale, the circumstellar envelope of 
V\,Hya is quasi-spherical. But in the heart of the nebula, the dust emission shows an extension along
 the East-West direction. This is clearer on the right panel of Fig. \ref{figvhya}, which shows the ellipticity
 of the observed circumstellar envelope as a function of the radius. Within 1'' to the star,
 the ellipticity steadily decrease from $\sim$1.2 to $\sim$1. V\,Hya thus displays a departure from spherical symmetry
 in its heart, like what is observed for the closest AGB star, IRC\,+10\,216 (Le\~ao et al. 2006). This elongation 
in the core of the
 envelope has the same orientation as the jet observed previously in CO. V\,Hya appears as an AGB star that was
 previously spherically symmetric, now being shaped by a high velocity jet. V\,Hya is thus an AGB star on his way 
to become an asymmetrical PPN and then PN. These observations thus show the importance of high speed jet for the
 formation of
 planetary nebula, as early as during the AGB phase.

\section{Conclusion}
The observation we presented here show that adaptive optics imaging imaging with NACO on the VLT is of great use to
 the study of small structures around evolved stars. We have shown that the shape of the nebula around Hen\,2-113
 could be explained by the projection of a diabolo-shaped structure and that hot dust was present close to its [WC10] central star. The NACO 2.12\,$\mu$m image of the PN Roberts\,22 revealed a
 very complex S-shaped structure embedded in a more diffuse rectangular-shaped envelope. Observations of these nebula
 at two different epochs seems to indicate that the S-shape  is in expansion at a speed of $\sim$ 450 km.s$^{-1}$. Our
 combined NACO/MIDI observations of OH\,231.8+4.2 have revealed the presence of a very compact flattened dusty structure
 in its core (40-50 a.u.). The shape of this structure is not the same at 8 and 13\,$\mu$m, due to emission 
from hot dust heated by the hot companion. Finally, the NACO image of the AGB star V\,Hya
 shows that its envelope is spherical at large scale and elongated toward a direction roughly East-West in its core.
 This asymmetry is probably due to a jet shaping the AGB circumstellar envelope that will become an asymmetric planetary
 nebula which should provide beautiful images for the conference APN\,5446 or more.

%
%
%
%
%
%



\end{document}